\documentclass{ws-procs961x669}            % default, citations in superscript
\begin{document}
\title{Some Recent Results on Ricci-Based Gravity Theories}

\author{Gonzalo J. Olmo$^{a,b}$ and Diego Rubiera-Garcia$^c$}

\address{$^a$ Departamento de F\'{i}sica Te\'{o}rica and IFIC, Centro Mixto Universidad de Valencia - CSIC.
Universidad de Valencia, Burjassot-46100, Valencia, Spain\\
E-mail: gonzalo.olmo@uv.es}
\address{$^b$ Universidade Federal do Cear\'a (UFC), Departamento de F\'isica,\\ Campus do Pici, Fortaleza - CE, C.P. 6030, 60455-760 - Brazil.}
%\author{}
\address{$^c$Departamento de F\'isica Te\'orica and IPARCOS,
	Universidad Complutense de Madrid, E-28040 Madrid, Spain\\
E-mail: drubiera@uv.es}
\begin{abstract}
Metric-affine theories in which the gravity Lagrangian is built using (projectively invariant) contractions of the Ricci tensor with itself and with the metric (Ricci-Based Gravity theories, or RBGs for short) are reviewed. The goal is to provide a contextualized and coherent presentation of some recent results. In particular, we focus on the correspondence that exists between the field equations of these  theories and those of general relativity, and comment on how this can be used to build new solutions of physical interest. We also discuss the formalism of junction conditions in the $f(R)$ case, and provide a brief summary on current experimental and observational bounds on model parameters. 
\end{abstract}

\keywords{metric-affine gravity}

\bodymatter

\section{Introduction}\label{sec:intro}

The discovery of the accelerated expansion of the universe by means of Type-IA supernovae in the last years of the past century stimulated a burst of theoretical activity aimed at explaining such an unexpected phenomenon\cite{Sn1,Sn2,Sn3}. Modifications of the matter sector via dark energy \cite{Peebles:2002gy}, and of the very equations that govern the cosmic dynamics according to the standard model defined by general relativity (GR) \cite{Joyce:2014kja}, were proposed as potential mechanisms to justify those and other observations regarding large scale structures, baryon acoustic oscillations, CMB, microlensing, etc \cite{P0,P1,P2,P3,BAO,DES,Bull:2015stt}, which were also compatible with an accelerating phase initiated at relatively recent cosmic times. Among the many proposals of that epoch, some of us paid attention to a family of modified gravity theories with slightly unusual properties, namely, Palatini (or metric-affine) $f(R)$ theories \cite{Olmo:2011uz}. Theories of the $f(R)$ type \cite{DeFelice:2010aj} arise as an attempt to encapsulate in a phenomenological approach various results coming from the theory of quantized fields in curved space-times and low-energy limits of string theories, which suggest that the action of GR, linear in the Ricci scalar $R$, must be supplemented with additional higher-order curvature invariants to account for new dynamics at high energies \cite{Parker:2009uva,Ortin:2015hya}. The field of modified gravity has considerably grown since then and the reader is invited to consult some of the excellent review articles and books available in the literature \cite{Capozziello:2011et,Clifton:2011jh,Nojiri:2017ncd,Heisenberg:2018vsk,Harko:2018ayt}.

Unlike other theories being considered in the literature at that time, Palatini $f(R)$ theories were able to produce modified dynamics without introducing new dynamical degrees of freedom.  This could happen in a non-perturbative way right at the {\it energy-density} scale (rather than just {\it energy} scale) defined by the parameters of the theory under consideration \cite{Barragan:2009sq,Barragan:2010qb}. Moreover, their field equations in vacuum turned out to recover exactly those of GR with an effective cosmological constant. Thus, these theories could be seen as some kind of {\it minimal extension} or deformation of GR induced by the presence of matter fields only at those energy-density scales, thus without any conflict with weak-field observations. In fact, this deformation can be associated to an auxiliary conformal geometry present in the theory which arises due to the connection field equation\footnote{ To understand this point, we note that, in the Palatini version of $f(R)$ theories, metric and affine connection are regarded as a priori independent geometric entities, both of which generate their own set of field equations upon the variational principle. For nonlinear $f(R)$ functions, the connection equation can be solved as the Levi-Civita connection of a metric $h_{\mu\nu}$ conformally related to the space-time metric as $g_{\mu\nu}=\frac{1}{f_R}h_{\mu\nu}$, with $f_R=df/dR$ being a function of the trace of the matter fields determined by the equation $R f_R-2f=\kappa^2 T$. In vacuum, $T=0$, one finds that $R=R_0=$constant satisfies  $R f_R-2f=0$, and implies that the conformal factor between metrics is just a constant, which amounts to a simple change of units. When $T\neq 0$, the deformation occurs in a continuous non-constant way. }. 

The exploration of $f(R)$ and of other extensions involving contractions of the Ricci tensor $R_{\mu\nu}$, such as $f(R,R_{\mu\nu}R^{\mu\nu})$, allowed to extend some results and generic properties beyond the $f(R)$ case. In particular, it was observed that Palatini $f(R,R_{\mu\nu}R^{\mu\nu})$ theories also lead to modified dynamics without introducing new degrees of freedom, that the field equations in vacuum coincide with those of GR+$\Lambda_{eff}$, and that the connection equation can be solved in terms of an auxiliary geometry which, nonetheless, is not conformal in general, but disformal. The consideration of other models involving other contractions of the Ricci tensor, such as those inspired by Born-Infeld theories \cite{Vollick,Banados} or written in terms of elementary symmetric polynomials \cite{BeltranJimenez:2014hma} confirmed this general trend as a robust and generic signature of Palatini theories constructed with the Ricci tensor. 

A subtle, though very important, aspect of these larger families of metric-affine theories is related to projective invariance. Projective transformations are a set of transformations of the connection of the form\cite{Eisenhart}
\begin{equation}
\tilde\Gamma^\mu_{\alpha\beta}=\Gamma^\mu_{\alpha\beta}+\xi_\alpha \delta^\mu_\beta \ ,
\end{equation}
such that the Riemann tensor, $ {R^\alpha}_{ \beta\mu\nu}=\partial_\mu\Gamma_{\nu\beta}^\alpha-\partial_\nu\Gamma_{\mu\beta}^\alpha+\Gamma_{\mu\lambda}^\alpha\Gamma_{\nu\beta}^\lambda-\Gamma_{\nu\lambda}^\alpha\Gamma_{\mu\beta}^\lambda$, transforms as 
\begin{equation}
\tilde {R^\alpha}_{\mu\beta\nu}={R^\alpha}_{\mu\beta\nu}+\delta^\alpha_\mu F_{\beta\nu} \ ,
\end{equation}
where $F_{\beta\nu}=\partial_\beta \xi_\nu-\partial_\nu \xi_\beta$. As one can see, the Ricci tensor becomes $\tilde R_{\mu\nu}=R_{\mu\nu}+F_{\mu\nu}$, and its symmetric part $\tilde R_{(\mu\nu)}=R_{(\mu\nu)}$ is invariant under projective transformations. Obviously, the Ricci scalar is also invariant under this type of transformations. The relevance of this is that though the field equations of metric-affine theories (with independent metric and connection) are generically of second order, one needs to impose projective invariance of the action in order to get rid of ghost-like dynamical degrees of freedom \cite{BeltranJimenez:2019acz}, which are dangerous for the stability of the theory \cite{BeltranJimenez:2020sqf} (see, however, Ref.\citenum{Deffayet:2021nnt}). Projective invariance is also very useful to solve the connection equation, as it allows to trivialize the role of torsion in many cases of interest \cite{Afonso:2017bxr}. All these results have allowed to identify a family of gravity theories, dubbed as Ricci-Based Gravity theories (RBGs), which are specified by a projective invariant action of the form\footnote{Note that we should now update the notation used in a previous paragraph to say $f(R,R_{(\mu\nu)}R^{(\mu\nu)})$ instead of $f(R,R_{\mu\nu}R^{\mu\nu})$ in order to be dealing with projectively invariant theories.}
\begin{equation}\label{eq:Action-0}
S_{RBG}(g_{\mu\nu},\Gamma^\alpha_{\beta\gamma},\Psi_m)=\frac{1}{2\kappa^2}\int d^4x \sqrt{-g} f(g^{\mu\alpha}R_{(\alpha\nu)})+S_m(g_{\mu\nu},\Psi_m) \ ,
\end{equation}
where the unspecified function $f$ depends on scalars constructed with traces of powers of the object $g^{\mu\alpha}R_{(\alpha\nu)}$, $S_m$ represents the matter action, and $\Psi_m$ denotes collectively the matter fields, which are assumed to be minimally coupled to the metric $g_{\mu\nu}$ \cite{Delhom:2020hkb}. 

In this paper we review recent progress made in the literature on RBGs trying to contextualize various technical achievements and applications, and also providing a coherent presentation of those results. In particular, we will focus on the description of a correspondence that allows to rewrite a given RBG model coupled to certain matter fields as a standard problem in GR but coupled to a modified matter source. This provides a powerful tool to deal with physically realistic scenarios which would require otherwise numerical methods for the resolution of the field equations. We will also comment on the problem of junction conditions and on the current constraints on model parameters. Since our purpose in mostly conceptual, we will keep the technical discussion to a minimum, providing references to the relevant literature when needed.

\section{Einstein frame representation}

The theory defined in (\ref{eq:Action-0}) can be manipulated to obtain a generic Einstein-frame representation, in which the gravitational sector is formally written like in GR. To do it, the simplest approach consists on introducing a set of auxiliary fields ${\Sigma^\mu}_\nu$ such that the action is rewritten as 
\begin{eqnarray}\label{eq:Action-1}
S_{RBG}(g_{\mu\nu},\Gamma^\alpha_{\beta\gamma},{\Sigma^\mu}_\nu,\Psi_m)&=&\frac{1}{2\kappa^2}\int d^4x \sqrt{-g}\left[ f({\Sigma^\mu}_\nu)+\left(g^{\mu\alpha}R_{(\alpha\nu)}-{\Sigma^\mu}_\nu\right)\frac{\partial f}{\partial {\Sigma^\mu}_\nu}\right]\nonumber \\ &+&S_m(g_{\mu\nu},\Psi_m) \ .
\end{eqnarray}
It is easy to see that variation with respect to ${\Sigma^\mu}_\nu$ leads to the conditions ${\Sigma^\mu}_\nu=g^{\mu\alpha}R_{(\alpha\nu)}$, subject to the constraint $\frac{\partial^2 f}{\partial {\Sigma^\mu}_\nu \partial {\Sigma^\rho}_\lambda}\neq 0$, such that when it is satisfied, the theory recovers the same classical solutions as (\ref{eq:Action-0}). Rearranging terms,  (\ref{eq:Action-1}) can also be written as 
\begin{eqnarray}\label{eq:Action-2}
S_{RBG}(g_{\mu\nu},\Gamma^\alpha_{\beta\gamma},{\Sigma^\mu}_\nu,\Psi_m)&=&\frac{1}{2\kappa^2}\int d^4x \sqrt{-g}\Big[g^{\mu\alpha}R_{(\alpha\nu)}\frac{\partial f}{\partial {\Sigma^\mu}_\nu} \nonumber \\
&+& \Big(f({\Sigma^\mu}_\nu)-{\Sigma^\mu}_\nu\frac{\partial f}{\partial {\Sigma^\mu}_\nu}\Big)\Big] \\ &+&S_m(g_{\mu\nu},\Psi_m) \ . \nonumber
\end{eqnarray}
The next step requires the introduction of an auxiliary metric $h_{\mu\nu}$ such that 
\begin{equation}
\sqrt{-h}h^{\mu\nu}\equiv \sqrt{-g}g^{\mu\alpha}\frac{\partial f}{\partial {\Sigma^\alpha}_\nu} \ ,
\end{equation}
which turns (\ref{eq:Action-2}) into 
\begin{equation}\label{eq:Action-3}
S_{RBG}(g_{\mu\nu},\Gamma^\alpha_{\beta\gamma},{\Sigma^\mu}_\nu,\Psi_m)=\frac{1}{2\kappa^2}\int d^4x \sqrt{-h}h^{\mu\nu}R_{(\mu\nu)} + \tilde{S}_m(g_{\mu\nu},{\Sigma^\alpha}_\beta,\Psi_m) \ ,
\end{equation}
where $\tilde{S}_m(g_{\mu\nu},{\Sigma^\alpha}_\beta,\Psi_m)$ now incorporates the other $\Sigma-$dependent terms in (\ref{eq:Action-2}). If one finds a way to rewrite this modified matter action in terms only of $h_{\mu\nu}$ and $\Psi_m$, then in the new variables the gravitational field equations of the original RBG theory minimally coupled to $S_m(g_{\mu\nu},\Psi_m)$ would look like the GR equations coupled to the stress-energy tensor derived from $\tilde S_m(h_{\mu\nu},\Psi_m)$, thus turning the original modified gravity problem into a standard GR problem. This task, however, is not a trivial one and for its efficient implementation it is usually more convenient to make use of the field equations and consider specific forms of matter fields that help obtain useful relations and simplifications that allow to establish the correspondence between the original RBG-frame variables and those in the Einstein frame. 

\section{Correspondence between RBG and GR variables}

The field equations of RBGs were derived in full detail in Ref.\citenum{Afonso:2017bxr}. Here, we just mention that for minimally coupled (and some non-minimally coupled) bosonic mater fields, and for fermions coupled via the Dirac equation (without explicitly including a coupling between the axial part of the torsion and the fermions), projective invariance allows to trivialize the role of torsion in the connection field equation, which can then be solved as the Levi-Civita connection of an auxiliary metric $h_{\mu\nu}$ related to $g_{\mu\nu}$ via a deformation matrix, namely,
\begin{equation}\label{eq:hgOmega}
h_{\mu\nu}=g_{\mu\alpha}{\Omega^\alpha}_\nu \ ,
\end{equation}
where ${\Omega^\alpha}_\nu={\Omega^\alpha}_\nu(T_{\rho\lambda})$ is a model-dependent function of the stress-energy tensor. It should be noted that, by the Cayley-Hamilton theorem and the theory of matrices,  the form of ${\Omega^\alpha}_\nu$ can be determined\footnote{In four dimensions.} as a sum involving the objects $\{\delta^\mu_\nu, {T^\mu}_\nu,  {T^\mu}_\alpha  {T^\alpha}_\nu, {T^\mu}_\alpha  {T^\alpha}_\beta {T^\beta}_\nu \}$ multiplied by suitable functions of the eigenvalues of ${T^\mu}_\nu$. This puts forward that, up to a trivial constant rescaling, the vacuum geometries defined by $h_{\mu\nu}$ and $g_{\mu\nu}$ are the same, experiencing departures only in regions containing matter fields.  It also means that the geometry is sensitive to the type and kind of matter fields present in each region. 

The metric field equations can be efficiently written in the Einstein-frame representation of the metric as
\begin{equation}\label{eq:EEmatrix}
{G^\mu}_\nu (h)= \frac{\kappa^2}{|\hat{\Omega}|^{1/2}} \left[{T^\mu}_\nu-\left(\mathcal{L}_G +\frac{T}{2}\right){\delta^\mu}_\nu\right]\ ,
\end{equation}
where ${G^\mu}_{\nu}(h) \equiv h^{\mu\alpha}G_{\alpha\nu}(h)=h^{\mu\alpha} (R_{\mu\nu}(h)-\frac{1}{2}h_{\mu\nu}R(h))$ is the Einstein tensor of the auxiliary metric $h_{\mu\nu}$, $\vert \hat{\Omega} \vert$ denotes the determinant of the matrix ${\Omega^\mu}_{\nu}$, $\mathcal{L}_G \equiv f/2\kappa^2$ is the gravity Lagrangian, and $T \equiv g^{\mu\nu}T_{\mu\nu}$ is the trace of the stress-energy tensor. Written in this form, it becomes apparent that the geometry associated to $h_{\mu\nu}$ is determined by the matter fields that represent the sources on the right-hand side of this equation. Thus, $h_{\mu\nu}$ is sensitive to the total amounts of matter and energy, whereas $g_{\mu\nu}$, as mentioned above, will also feel the presence of the fields stress-energy densities. 

An obvious advantage of writing the metric field equations in terms of $h_{\mu\nu}$ is that the left-hand side of (\ref{eq:EEmatrix}) adopts the usual Einstein form, thus showing that the metric satisfies a set of second-order field equations for which there exist different methods to obtain solutions. Moreover, it is evident that in vacuum the equations are just those of GR \cite{Ferraris:1992dx,Borowiec:1996kg}, with only two degrees of freedom that propagate at the speed of light, thus being compatible with current observations\footnote{See \cite{Jana:2017ost} for a discussion on this point using Eddington-inspired Born-Infeld gravity, a member of the RBG class.}.  A disadvantage is that its right-hand side is written in terms of $T_{\mu\nu}=-(2/\sqrt{-g})\delta S_m/\delta g^{\mu\nu}$ and, therefore, it depends on the metric $g_{\mu\nu}$, whose relation with $h_{\mu\nu}$ is not always obvious or does not allow for a simple translation. Thus, the technical challenge consists on defining algorithms or strategies that allow to express the right-hand side of that equation as the stress-energy tensor of matter fields minimally coupled to the metric $h_{\mu\nu}$, such that one would have ${G^\mu}_\nu (h)= {\kappa^2}{\tilde T^\mu}_\nu$, with $\tilde T_{\mu\nu}=-(2/\sqrt{-h})\delta \tilde S_m/\delta h^{\mu\nu}$. 

\subsection{Scalar fields}

To illustrate the general comments of above, let us consider a real scalar field with an action of the form
\begin{equation} \label{eq:scalarRBG}
{S}_m(X,\phi)=-\frac{1}{2}\int d^4x\sqrt{-g}P(X,\phi) \ ,
\end{equation}
where $X=g^{\alpha\beta}\partial_\alpha\phi\partial_\beta\phi$ is the trace of ${X^\mu}_\nu\equiv g^{\mu\alpha}\partial_\alpha\phi\partial_\nu\phi$ and $P$ is some arbitrary function of its arguments. The associated stress-energy tensor can be written as 
\begin{equation}\label{eq:TmnX}
{{T}^\mu}_\nu=P_X {X^\mu}_\nu- \frac{P(X,\phi)}{2}{\delta^\mu}_\nu \ ,
\end{equation}
where $P_X \equiv dP/dX$. Given this structure and having in mind the Cayley-Hamilton theorem mentioned above, one finds that ${\Omega^\mu}_\nu$ must  have the form
\begin{equation}\label{eq:OmegaX}
{\Omega^\mu}_\nu=C(X,\phi){\delta^\mu}_{\nu}+D(X,\phi){X^\mu}_\nu \ ,
\end{equation}
%disformal transformation?
 where $C(X,\phi)$ and $D(X,\phi)$ are model-dependent functions. Using this result in Eq.(\ref{eq:hgOmega}), one finds that  
\begin{equation}\label{eq:Xmn2Zmn}
{X^\mu}_\nu=(C+DX){Z^\mu}_\nu \  \Rightarrow \ Z=\frac{X}{C+DX} \ ,
\end{equation}
where ${Z^\mu}_\nu=h^{\mu\alpha}\partial_\alpha\phi \partial_\nu\phi$ is the kinetic term minimally coupled to $h_{\mu\nu}$ that one uses in the construction of the $\tilde S_m$ action
\begin{equation} \label{eq:scalarGR}
\tilde{{S}}_m(Z,\phi)=-\frac{1}{2}\int d^4x\sqrt{-q}K(Z,\phi) \ .
\end{equation}
From (\ref{eq:Xmn2Zmn}) one sees that $Z=Z(X,\phi)$ can, in principle,  be used to obtain an expression for $X=X(Z,\phi)$, thus implying that the right-hand-side of Eq.(\ref{eq:EEmatrix}) can be written as the stress-energy tensor of the scalar field theory defined by $\tilde{{S}}_m(Z,\phi)$.

In order to establish the correspondence between the RBG theory coupled to the scalar matter action (\ref{eq:scalarRBG}) and GR coupled to (\ref{eq:scalarGR}), it is necessary to solve the relation\cite{Afonso:2018hyj}
\begin{eqnarray}\label{eq:themap}
{\tilde{T}^\mu}_{\ \ \nu} & =& K_Z {Z^\mu}_\nu- \frac{K(Z,\phi)}{2}{\delta^\mu}_\nu \\
&=& \frac{1}{|\hat{\Omega}|^{1/2}} \left[P_X{X^\mu}_\nu-\left(\mathcal{L}_G +\frac{X P_X-P}{2}\right){\delta^\mu}_\nu\right] \ , \nonumber
\end{eqnarray}
in a way that be consistent with the evolution equation of the scalar field written in the two frames, namely
\begin{equation}
\partial_\mu\left(\sqrt{-g}P_Xg^{\mu\alpha}\partial_\alpha\phi\right)-\sqrt{-g}\frac{P_\phi}{2}=0 \ ,
\end{equation}
and
\begin{equation}
\partial_\mu\left(\sqrt{-q}K_Zq^{\mu\alpha}\partial_\alpha\phi\right)-\sqrt{-q}\frac{K_\phi}{2}=0 \ .
\end{equation}
With a bit of algebra, one eventually finds that the $K(Z,\phi)$ Lagrangian can be written as 
\begin{equation}
K(Z,\phi)= \frac{1}{|\hat{\Omega}|^{1/2}}\left(2\mathcal{L}_G +{X P_X-P}\right) \label{eq:diag} \ ,
\end{equation}
which should be suplemented by  (\ref{eq:Xmn2Zmn}) in order to obtain $X=X(Z,\phi)$. This shows that given an RBG theory coupled to a scalar field source of the form (\ref{eq:scalarRBG}), it is possible to explicitly construct an Einstein frame representation in which the scalar field is minimally coupled to the Einstein frame metric $h_{\mu\nu}$ via the Lagrangian $K(Z,\phi)$ defined in (\ref{eq:diag}). 

The summary presented here can be found in full detail in Ref.\citenum{Afonso:2018hyj}, where the analysis is extended also to theories with an arbitrary number of scalar fields. Applications of this approach have also been considered in several works. In Ref.\citenum{Afonso:2019fzv} it was shown that correspondence between theories can also be considered in the reversed way, namely, from GR plus matter to RBGs with a different matter Lagrangian. In particular, starting with the Janis-Newman-Winicour free scalar field solution of GR \cite{JNW:1968}, new compact objects solutions where constructed analytically in two target theories: quadratic $f(R)$ and the Eddington-inspired Born-Infeld (EiBI) model. Interestingly, those theories exhibit solutions that behave like wormholes but which, nonetheless, are geodesically incomplete despite the finiteness of curvature invariants everywhere. More recently, in Ref.\citenum{Renan2022} a similar analysis was performed for arbitrary $f(R, R_{(\mu\nu)} R^{(\mu\nu)})$ Lagrangian, finding again geodesically incomplete wormhole solutions but identifying geometric properties that may be common to theories with different couplings between geometry and scalar fields. Rotating solutions involving a free scalar field in GR mapped into the EiBI theory have also been obtained \cite{Shao:2020weq}. 

Aside from those analytical examples, which provide a proof of concept for the mapping method, numerical studies of boson stars in Palatini $f(R)$ theories have also been worked out. In particular, in Ref.\citenum{Maso-Ferrando:2021ngp} the correspondence described above was used to find equilibrium solutions of boson stars involving a canonical massive complex scalar field coupled to $f(R)=R+\xi R^2$. The mapping to the GR frame implies that the transformed scalar becomes a non-canonical field with a non-standard coupling between the kinetic term and the potential. To be precise, if $P(X,\Phi)=X-2V$ (with $X=g^{\mu\nu}\partial_\mu \bar{\Phi}\partial_\nu {\Phi}$ and $V=-\mu^2\bar{\Phi}\Phi$), then the mapping implies 
\begin{equation}
K(Z,\Phi)=\frac{Z-\xi \kappa^2 Z^2}{1-8\xi \kappa^2 V}-\frac{2V}{1-8\xi \kappa^2 V} \ .
\end{equation}
One then solves the GR plus exotic matter problem first and then uses the results to generate the solution in the original frame. This approach has the advantage that one ends up solving two different problems, which may have phenomenological interest on their own. In this particular case, one observes that there exists a strong degeneracy between the observable properties of the $f(R)$ plus canonical scalar and the GR plus exotic scalar. This degeneracy worsens if self-interactions are added in the potential.

\subsection{Fluids}

If one considers a fluid as the matter source, the most general situation is that in which the stress-energy tensor can be diagonalized and has four different eigenvalues. Using an orthonormal frame for illustrative purposes, the stress-energy tensor of the fluid can be expressed as 
\begin{equation}\label{eq:T3ori}
T_{\mu\nu}=\rho u_\mu u_\nu+\sum_{i=1}^{i=3} P_i \xi^{(i)}_\mu \xi^{(i)}_\nu \ ,
\end{equation}
where $\rho$ represents the energy density and the $P_i$ are the main pressures, while $u^\mu$ is a normalized time-like vector and the $\xi^{(i)}_\mu$ represent an orthogonal spatial basis. Given the completeness relation $\delta^\mu_\nu=-u^\mu u_\nu+\sum_{i=1}^{i=3} \xi^{(i)\mu} \xi^{(i)}_\nu$, one can replace one of the projectors by the others plus the identity, such that $\xi^{(3)\mu} \xi^{(3)}_\nu=\delta^\mu_\nu+u^\mu u_\nu-\xi^{(1)\mu} \xi^{(1)}_\nu-\xi^{(2)\mu}\xi^{(2)}_\nu$, turning (\ref{eq:T3ori}) into
\begin{equation}\label{eq:T3}
{T^\mu}_\nu=(\rho+P_3) u^\mu u_\nu+(P_1-P_3)\xi^{(1)\mu} \xi^{(1)}_\nu+(P_2-P_3)\xi^{(2)\mu}\xi^{(2)}_\nu+P_3\delta^\mu_\nu \ .
\end{equation}
From this expression it is easy to see what happens when degeneracies are present. If, for instance, one assumes that $P_2=P_3$, then we obtain 
\begin{equation}\label{eq:T2}
{T^\mu}_\nu=(\rho+P_3) u^\mu u_\nu+(P_1-P_3)\xi^{(1)\mu}\xi^{(1)}_\nu+P_3\delta^\mu_\nu \ .
\end{equation}
And when the three pressures coincide, we recover the usual expression for a perfect fluid, namely, 
\begin{equation}\label{eq:T3}
{T^\mu}_\nu=(\rho+P) u^\mu u_\nu+P\delta^\mu_\nu \ .
\end{equation}
This fluid representation is very useful and versatile, allowing to provide a simple description of the mapping for different kinds of sources and combinations of them. In order to make contact with the existing literature, we will focus on the anisotropic fluid case of Eq.(\ref{eq:T2}), which we rewrite as 
\begin{equation}\label{eq:T2b}
{T^\mu}_\nu=(\rho+P_{\perp}) u^\mu u_\nu+(P_r-P_{\perp})\chi^\mu \chi_\nu+P_{\perp}\delta^\mu_\nu \ .
\end{equation}
to lighten the notation. Here we have $g_{\mu\nu}u^{\mu}u^{\nu}=-1$, $g_{\mu\nu}\chi^{\mu}\chi^{\nu}=+1$, $\rho$ is the fluid energy density, $p_r$ its pressure in the direction of $\chi^{\mu}$, and $p_{\perp}$ its tangential pressure in the direction orthogonal to $\chi^{\mu}$. Due to the orthogonality of the vectors involved, the structure of the deformation matrix must be
\begin{equation}  \label{eq:Omegafluid}
{\Omega^\mu}_{\nu}=\alpha {\delta^\mu}_{\nu} + \beta u^{\mu}u_{\nu} + \gamma \chi^{\mu}\chi_{\nu} \ ,
\end{equation}
where the explicit expressions of  $\{\alpha,\beta,\gamma\}$ are model dependent. Introducing these expressions into the RBG field equations (\ref{eq:EEmatrix}) we get
\begin{equation}
{G^\mu}_{\nu}(h)=\frac{\kappa^2}{\vert \hat{\Omega}\vert^{1/2}} \Big[\Big(\frac{\rho-p_r}{2}-\mathcal{L}_G\Big){\delta^\mu}_{\nu} +(\rho+p_{\perp})u^{\mu}u_{\nu} +(p_r-p_{\perp}) \chi^{\mu}\chi_{\nu} \Big] \ .
\end{equation}
Proposing another anisotropic fluid on the GR side, with new functions $\{\rho^h,p_r^h,p_{\perp}^h\}$ and orthonormal vectors $h_{\mu\nu}v^{\mu}v^{\nu}=-1$ and $h_{\mu\nu}\xi^{\mu}\xi^{\nu}=+1$, then the mapping equations become
\begin{eqnarray}
p_{\perp}^h&=&\frac{1}{\vert \hat{\Omega} \vert^{1/2}} \left[\frac{\rho-p_r}{2} -\mathcal{L}_G \right] \label{eq:mapflu1} \\
\rho^h+p_{\perp}^h&=&\frac{\rho+p_{\perp}}{\vert \hat{\Omega} \vert^{1/2}}  \label{eq:mapflu2} \\
p_r^h-p_{\perp}^h&=&\frac{p_r-p_{\perp}}{\vert \hat{\Omega} \vert^{1/2}}  \ . \label{eq:mapflu3}
\end{eqnarray}
These equations describe the correspondence between the two sets of scalars $\{\rho,p_r,p_{\perp}\}$ and $\{\rho^h,p_r^h,p_{\perp}^h\}$ once the RBG Lagrangian $\mathcal{L}_G$ is given. Together with the relations $u^{\mu}u_{\nu}=v^{\mu}v_{\nu}$ and $\chi^{\mu}\chi_{\nu} =\xi^{\mu}\xi_{\nu}$, this allows to write ${\Omega^\mu}_{\nu}$ in Eq.(\ref{eq:Omegafluid}) in terms of  the solution obtained in GR. This means that one can also find an expression for  ${\Omega^\mu}_{\nu}$ in the form
\begin{equation}  \label{eq:Omegafluid2}
{\Omega^\mu}_{\nu}=\tilde\alpha {\delta^\mu}_{\nu} +\tilde \beta v^{\mu}v_{\nu} + \tilde\gamma \xi^{\mu}\xi_{\nu} \ ,
\end{equation}
which allows to generate solutions of the RBG starting from a given seed solution of GR.

In order to provide an explicit example, let us consider EiBI gravity, whose action can be expressed as \cite{BeltranJimenez:2017doy}
\begin{equation} \label{eq:actionEiBI} %$\Lambda_{eff}=\frac{\lambda-1}{\kappa^2 \epsilon}$
\mathcal{S}_{EiBI}= \frac{1}{\kappa^2 \epsilon} \int d^4x \left(\sqrt{-h} - \lambda \sqrt{-g}\right) \ ,
\end{equation}
where $h_{\mu\nu} \equiv g_{\mu\nu} + \epsilon R_{\mu\nu}(\Gamma)$ denotes the connection-compatible metric, and $\epsilon$ is a parameter with dimensions of length squared.  For the action (\ref{eq:actionEiBI}) the deformation matrix is given by  \cite{BeltranJimenez:2017doy}
\begin{equation} \label{eq:OmegaEiBI}
\vert \hat{\Omega} \vert^{1/2} {(\Omega^{-1})^\mu}_{\nu}=\lambda {\delta^\mu}_{\nu} -\kappa^2 \epsilon {T^\mu}_{\nu} \ ,
\end{equation}
while the EiBI gravity Lagrangian $\mathcal{L}_G$ can be conveniently expressed as
\begin{equation} \label{eq:LEiBIos}
\mathcal{L}_G=\frac{\vert \hat{\Omega} \vert^{1/2}-\lambda}{\kappa^2\epsilon} \ .
\end{equation}
Using Eq.(\ref{eq:OmegaEiBI}) and the stress-energy tensor of the anisotropic fluid (\ref{eq:T2b}), one finds
\begin{equation} \label{eq:OmegaEiBI_JF}
\vert \hat{\Omega} \vert^{1/2} ({\Omega^\mu}_{\nu})^{-1}=(\lambda-\epsilon\kappa^2p_{\perp}) {\delta^\mu}_{\nu}  -\kappa^2 \epsilon \left[(\rho+p_{\perp})u^\mu u_\nu+(p_r-p_{\perp})\chi^\mu \chi_\nu\right] \ , 
\end{equation}
which can also be written in terms of the Einstein frame variables  as
\begin{equation} \label{eq:OmegaEiBI_EF}
({\Omega^\mu}_{\nu})^{-1}=\left(1-\frac{\epsilon\kappa^2}{2}[\rho^h-p^h_r]\right) {\delta^\mu}_{\nu}  -\kappa^2 \epsilon \left[(\rho^h+p_{\perp}^h)v^\mu v_\nu+(p^h_r-p^h_{\perp})\xi^\mu \xi_\nu\right]\ .
\end{equation}
The space-time metric in the RBG frame can thus be written using the Einstein frame variables as
\begin{equation} \label{eq:gqEiBI}
g_{\mu\nu}=\left(1-\frac{\epsilon\kappa^2}{2}[\rho^h-p^h_r]\right) h_{\mu\nu}  - \kappa^2 \epsilon \left[(\rho^h+p_{\perp}^h)v_\mu v_\nu+(p^h_r-p^h_{\perp})\xi_\mu \xi_\nu\right]\ . 
\end{equation}
This last relation provides a solution for the EiBI theory starting from any known seed solution in GR supported by an anisotropic fluid source. 

The first application of this kind was carried out in Ref.\citenum{Afonso:2018mxn}, where it was shown that the static, spherically symmetric solutions of EiBI gravity coupled to Maxwell's electrodynamics could be generated using as seed solution the corresponding one in GR coupled to the Born-Infeld nonlinear electrodynamics theory, clarifying in this way a puzzle that had emerge already in Ref.\cite{Olmo:2012nx} regarding the similarity between the solutions of quadratic $f(R, R_{(\mu\nu)}R^{(\mu\nu)})$ gravity coupled to Maxwell electrodynamics and the Born-Infeld case in GR. This result was further reinforced soon after in Ref.\citenum{Delhom:2019zrb}, where the correspondence between EiBI+Maxwell and GR+Born-Infeld was shown to be exact regardless of the symmetries of the particular scenario considered. This result was obtained working out the correspondence directly with electromagnetic fields, without using the fluid analogy. Having established these solid results, the first exact rotating solution in EiBI theory was derived in Ref.\citenum{Guerrero:2020azx}, where the Kerr-Newman solution of GR acted as a seed to generate the solution of EiBI coupled to Born-Infeld electrodynamics. In this line, the electromagnetic correspondence also allowed to obtain the first example of a multicenter configuration \cite{Olmo:2020fnk} in the framework of EiBI gravity. It turns out that this solution actually represents a set of wormholes in equilibrium rather than a set of point-like objects in equilibrium, as is the case in GR. Unfortunately, these objects exhibit curvature divergences and some other undesired features, which makes such solutions to have  little physical interest. The much more appealing case of finding multicenter solutions in EiBI gravity coupled to Maxwell electrodynamics thus remains as an open question. Rotating solutions in $2+1$ dimensions in EiBI gravity have also been obtained using the fluid and electromagnetic forms of the mapping \cite{Guerrero:2021avm}, providing in this way a double consistency check to this way of finding solutions in RBGs starting from known solutions in GR. The greatest achievement so far of the correspondence using fluids has been reported in Ref.\citenum{Afonso:2021pga}, where an infinite class of axially symmetric rotating solutions has been explicitly constructed for the EiBI theory, though the same logic can be applied to any other RBG. Hopefully, this result will allow to generate a catalogue of new rotating solutions that can be used to confront modified gravity effects with observational data.

\section{Junction conditions in Palatini $f(R)$ gravity and applications}

A relevant question that is not always properly addressed in the literature has to do with the presence of surfaces or boundaries of discontinuity, such as in scenarios that match two different bulk solutions across a given hypersurface \cite{Israel:1966rt,Darmois}. This happens, for instance in stellar models that are matched to an external vacuum geometry, in certain types of wormhole solutions constructed via a cut-and-paste procedure, in braneworld models in which the matter fields are constrained to live on a four-dimensional hypersurface, in cosmic string/domain wall configurations, etc. In all those cases, the domain of definition of certain tensorial quantities have a compact support, and the field equations must be handled with care in order to be mathematically consistent with the theory of tensorial distributions. In this sense, the first attempt to provide a rigorous description of the junction conditions in Palatini $f(R)$ theories was considered in Ref.\citenum{Olmo:2020fri}. Given that the vacuum solutions of these theories recover the equations of GR but within the sources they develop modified dynamics, it was natural to expect some modification of the junction conditions as compared to those corresponding to GR. And this is indeed what happens. 

From an intuitive perspective based on the covariant description of the bulk equations, the fact that the connection is related to a conformal geometry whose conformal factor depends on the trace of the stress-energy tensor, $T$, suggests that in order to have well-defined geometries one should have continuous derivatives of the conformal factor (hence of $T$) up to second order. However, the distributional analysis puts forward that only $T$ must be continuous, thus allowing discontinuities in its derivative across the matching surface. At the same time, the singular part of the stress-energy tensor (living on the hypersurface) should also be traceless (see Ref.\citenum{Olmo:2020fri} for details). This last result contrasts with what happens in GR and in the metric version of $f(R)$ theories \cite{Senovilla:2013vra,Deruelle:2007pt,Padilla:2012ze,Vignolo:2018eco,delaCruz-Dombriz:2014zaa}, where this trace represents the {\it brane tension} and nothing forces it to be vanishing. Finally, the discontinuity in the second fundamental form $K^\pm_{\mu\nu}\equiv {h^\rho}_\mu {h^\sigma}_\nu \nabla_\rho^\pm n_\sigma$ (see Ref.\citenum{Olmo:2020fri}  for notation and conventions) is given by 
\begin{equation}\label{eq:fR}
-[K_{\mu\nu}]+\frac{1}{3}h_{\mu\nu} [K^\rho_\rho]=\kappa^2\frac{\tau_{\mu\nu}}{f_{R_\Sigma}} \ ,
\end{equation}
which contrasts with the Israel-Darmois \cite{Israel:1966rt,Darmois} GR equation
\begin{equation}\label{eq:GR}
-[K_{\mu\nu}]+h_{\mu\nu} [K^\rho_\rho]=\kappa^2{\tau_{\mu\nu}} \ .
\end{equation}
Note that, as mentioned above, the trace of (\ref{eq:fR}) vanishes, thus implying certain constraints on the kind of sources that can exist on the boundary layer. Regarding the Bianchi identities, one finds that the energy conservation equation remains the same as in GR, namely, $D^\rho \tau_{\rho\nu}=-n^\rho {h^\sigma}_\nu [T_{\rho\sigma}]$, but the second condition becomes 
\begin{equation}
(K^+_{\rho\sigma}+K^-_{\rho\sigma})\tau^{\rho\sigma}=2 n^\rho n^\sigma [T_{\rho\sigma}]-\frac{3R_T^2f_{RR}^2}{f_R}[b^2] \ ,
\end{equation}
where $[b^2]\equiv [(n^\alpha \nabla_\alpha T)^2]$. Note that this equation smoothly recovers the GR result when $f_{RR}\to 0$. 

When the above equations are applied to stellar models involving a polytropic fluid matched to an exterior vacuum solution, one finds that the range of polytropic indices which can be considered without running into trouble with the appearance of curvature divergences at the shell improves as compared to a naive approach based only on the bulk equations (see Refs.\citenum{Barausse:2007pn,Barausse:2007ys,Barausse:2008nm} for a first analysis in this regard and Ref.\citenum{Olmo:2008pv} for the subsequent discussion). Thus, this confirms that neutron stars and white dwarfs can be safely modeled within the Palatini $f(R)$ framework. 

The modified junction conditions equations derived above have already been used to study the existence and stability of thin-shell wormholes \cite{Lobo:2020vqh}. Unlike in GR, one finds that the number of degrees of freedom on the shell is reduced to a single one due to the constraint $\tau=0$ on the trace of the singular part of the stress-energy tensor, which establishes a relation between the surface energy density and the surface tension. As a result, this modifies the dynamical equations as compared to GR with the peculiarity that the resulting equations are the same for any nonlinear Palatini $f(R)$ theory. When applied to thin-shell wormholes constructed by a cut-and-paste procedure using Schwarzschild or Reissner-Nordstrom geometries, one finds that the Schwarzschild case is always unstable, whereas stable solutions exist in the Reissner-Nordstrom case, including reflection asymmetric scenarios \cite{Guerrero:2021pxt}. Moreover, unlike in GR, the stability condition can be achieved even for positive energy-density configurations.

\section{Observational constraints}

In addition to the technical advances carried out in the last years in order to better understand the structure and dynamics of Palatini $f(R)$ theories and other RBGs, substantial progress has been achieved also in order to constrain model parameters using astrophysical and laboratory data. In this regard, for the leading-order quadratic corrections, $f(R)=R+\xi R^2$, an analysis of the weak-field limit \cite{Olmo:2005zr,Olmo:2005hc} leads to $|\xi|\ll 2\times 10^{12}$ cm$^2$. From heuristic considerations based on nuclear physics \cite{Avelino:2012qe,BeltranJimenez:2017doy}, one concludes that  $|\xi|< 6\times 10^{9}$ cm$^2$ (equivalently, $\rho_\xi\equiv 1/(\kappa^2\xi)\ge 10^{22}$ g/cm$^3$). By looking at the effects that the modified Poisson equation implies for brown dwarf stars \cite{Olmo:2019qsj,Olmo:2019flu,Olmo:2021yul,CANTATA:2021ktz}, it is also possible to conclude that $\rho_\xi> 10^{5}$ g/cm$^3$, which is not a strong bound but could improve in the next few years with increasing statistics of the brown dwarf population and their mass spectrum.  The strongest bounds so far on generic quadratic corrections come from big bang nucleosynthesis data and from the compatibility of the RBG dynamics with elementary particle experiments. In particular, current bounds on light-by-light and electron-electron scattering set a limit  on generic quadratic corrections of order $\rho_\xi\ge 10^{25}$ g/cm$^3$ \cite{Jimenez:2021sqd,Latorre:2017uve,Delhom:2019wir}, while  compatibility of current cosmological data for cosmic chronometers, standard candles, BAO, and CMB with the predictions of the EiBI model \cite{Benisty:2021laq} imply that $\rho_\epsilon \ge 10^{35}$ g/cm$^3$. This clearly constrains the domain of modified gravity effects to very early epochs of the cosmic expansion and to extremely high energy densities.

\section{Conclusion}

In this work we have briefly reviewed some recent progress made within the framework of Ricci-Based Gravity theories in the metric-affine formulation. We have shown that the space of solutions of these theories can be put into correspondence with the space of solutions of GR by means of algebraic manipulations and field redefinitions. This offers a clear practical advantage in the search of solutions, as one can now study modified gravity problems within GR itself (modified gravity without modified gravity). In particular, the full weaponry of analytical and numerical methods available for GR is now also available to deal with this family of gravity theories, thus saving a substantial amount of human and technical resources. We have also seen that a proper treatment of the junction conditions confirms that polytropic stellar models of physical interest are free of pathologies when matched to an exterior vacuum solution, which clarifies reasonable doubts that emerge when only the bulk equations are used to build solutions with a matching surface. Finally, a brief discussion of current bounds on model parameters from laboratory and astrophysical considerations has been provided. 
Before concluding, we must say that RBGs are just a portion of a larger family of metric-affine theories of gravity. Though much more room exists to include propagating degrees of freedom associated to torsion and non-metricity, one should not forget that any such theory must contain an RBG part, which makes relevant all the methods and results presented here. 

\section*{Acknowledgments}

 DRG is funded by the {\it Atracci\'on de Talento Investigador} program of the Comunidad de Madrid (Spain) No. 2018-T1/TIC-10431. This work is supported by the Spanish Grants FIS2017-84440-C2-1-P, PID2019-108485GB-I00, PID2020-116567GB-C21 and PID2020-116567GB-C21 funded by MCIN/AEI/10.13039/501100011033 (``ERDF A way of making Europe" and ``PGC Generaci\'on de Conocimiento"), the project PROMETEO/2020/079 (Generalitat Valenciana), the project H2020-MSCA-RISE-2017 Grant FunFiCO- 777740, the project i-COOPB20462 (CSIC),  the FCT projects No. PTDC/FIS-PAR/31938/2017 and PTDC/FIS-OUT/29048/2017, and the Edital 006/2018 PRONEX (FAPESQ-PB/CNPQ, Brazil, Grant 0015/2019). 

\bibliographystyle{ws-procs961x669}
\bibliography{ws-pro-sample}

%Non BiBTeX users can list down their references as:

\end{document}